\documentclass[twocolumn,aps,prb,showpacs,amsmath,superscriptaddress]{revtex4}
\usepackage{amssymb}
\usepackage{mathrsfs}
\usepackage{graphicx}
\usepackage{float}

\begin{document}
\title{Topological Mott insulators of ultracold atomic mixtures induced by interactions in one-dimensional optical superlattices}
\author{Zhihao Xu}
\affiliation{Beijing National Laboratory for Condensed Matter
Physics, Institute of Physics, Chinese Academy of Sciences,
Beijing 100190, China}
\author{Shu Chen}
\email{schen@aphy.iphy.ac.cn} \affiliation{Beijing National
Laboratory for Condensed Matter Physics, Institute of Physics,
Chinese Academy of Sciences, Beijing 100190, China}
\date{\today}

\begin{abstract}
We present exactly solvable examples that topological Mott
insulators can emerge from topologically trivial states due to
strong interactions between atoms for atomic mixtures trapped in
one-dimensional optical superlattice systems. The topological Mott
insulating state is characterized by nonzero Chern number and
appears in the strongly interacting limit as long as the total
band filling factor is an integer, which is not sensitive to the
filling of each component. The topological nature of the Mott
phase can be revealed by observing the density profile of the
trapped system. Our results can be also generalized to the multi-component atomic systems.

\end{abstract}

\maketitle

\section{ Introduction}
Topological band insulators have attracted
great attention in condensed matter physics since they were
discovered in spin-orbit coupled two-and three-dimensional systems
\cite{Kane,Qi}. As the topological insulator (TI) is expected to
be robust against weak perturbations, it is important to consider
the effect of interactions on the topological band insulators
\cite{Dzero,WangZ,Yu,Shen}. Although generally a strong
interaction may open the energy gap and break the TI, there exists
a class of topological insulator called as the topological Mott
insulator (TMI), where the interaction effects are responsible for
TI behavior \cite{Raghu}. The topological Mott phases have
attracted increasing attention \cite{Pesin,Ran,Zhang,TMI1,TMI2,TMI3,TMI4,TMI5} since the
concept is proposed as interaction plays a crucial role in the
formation of both the Mott phase and nontrivial topology. As
almost all these results based on the mean-field approximation,
examples with the interaction effect counted exactly is
particularly important for our understanding of the TMI.


In this work, we explore the realization of topological Mott
phases in cold atomic systems trapped in one-dimensional (1D)
quasi-periodic optical lattices, which can be generated by
superimposing two 1D optical lattices with commensurate or
incommensurate wavelengths \cite{Fallani,Roati,Deissler}. Cold
atomic systems in 1D quasi-periodic lattices have been extensively
studied \cite{Roscilde,Cai,Giamarchi,Yamashita} with a focus on
the Anderson localization \cite{Roati}. However, their nontrivial
topological features are recognized only very recently
\cite{Lang,Kraus}. Particularly, with the experimental observation
of the topological edge states in 1D photonic quasi-crystals
\cite{Kraus} there is a growing interest in the study of
topological properties in the 1D quasi-periodic lattices
\cite{Mei,Xu,Tezuka,Lang2,Viyuela,Kraus2}. It has been shown that
the free fermion system with its sub-bands being fully filled is a
topological nontrivial insulator characterized by a nonzero Chern
number in a two-dimensional (2D) parameter space \cite{Lang}. In
this work, we study the interacting atomic
mixtures in the 1D superlattice with its sub-bands are partially
filled by atoms. In the absence of inter-component interaction,
the system may be a Fermi metal, a Bose superfluid or their
mixture, depending on the component of mixture being fermion or
boson. We find that a TMI may emerge in the strongly interacting
regime if the total band filling is an integer. This conclusion is
exact in the limit of infinitely strong interaction (ISI) as it is
based on an exact mapping which relates the many-body wavefunction
of two-component mixture to the wavefunction of free fermion
system. Our study provides a simple way to realize the TMI in cold
atomic systems and may deepen our understanding of the TMI.


\section{Models and results}

\subsection{Models of interacting atomic mixtures in optical superlattices}
We consider the 1D atomic mixtures loaded in a bichromatic optical
lattice \cite{Fallani,Roati,Deissler}, which is described by
$H=H_0 + H_I$ with
\begin{equation}
H_0= -t \sum_{i,\sigma=\uparrow,\downarrow}
(\hat{c}^\dagger_{i,\sigma} \hat{c}_{i+1,\sigma}+ \mathrm{H.c.}) +
\sum_{i,\sigma=\uparrow,\downarrow} V_i \hat{n}_{i,\sigma}
\label{H0}
\end{equation}
and
\begin{equation}
H_I = U \sum_i \hat{n}_{i,\uparrow} \hat{n}_{i,\downarrow} +
\sum_{i} \frac 12 U_{\sigma} \hat{n}_{i,\sigma} \left(
\hat{n}_{i,\sigma}-1\right), \label{HI}
\end{equation}
where $V_i=\lambda \mathrm{cos}(2\pi \alpha i+\delta)$ with
$\lambda$ controlling the strength of commensurate potential,
$\alpha$ tuning the modulation period and $\delta$ being an
arbitrary phase, $c_{i,\sigma }$ are bosonic or fermionic
annihilation operators localized on site $i$, and $n_{i\sigma
}=c_{i_\sigma }^{\dagger }c_{i_\sigma }$. Here
$\sigma=\uparrow,\downarrow$ denotes the pseudo-spin index of two
components of atomic mixtures, which can be either fermion or
boson. For the two-component systems, there are three kinds of
mixtures, i.e., Fermi-Fermi (FF) mixture, Bose-Bose (BB) mixture and
Fermi-Bose (FB) mixture.
The interaction parameter $U$ denotes the inter-component
interaction strength, and $U_{\sigma}$ the intra-component
interaction strength with $U_{\sigma}=0$ between fermionic atoms.
Both $U$ and
$U_{\sigma}$ can be experimentally tuned to the limit of ISI by the Feshbach
resonance \cite{Feshbach}. The hopping amplitude $t$ is set to be the energy unit $(t=1)$.

In the absence of interactions, the eigenvalue equation, namely
the Harper equation, is given by
\begin{equation}
 -[\phi_{n}(i+1)+\phi_{n}(i-1)]+\lambda \mathrm{cos}(2\pi \alpha
i+\delta) \phi_{n}(i)=\epsilon_n \phi_{n}(i), \label{Harper1}
\end{equation}
where $\phi_{n}(i)$ denotes the single particle wave function and
$\epsilon_n$ the $n$-th single particle eigenenergy \cite{Hofstadter,Doh,Supp1}. For a
superlattice with $\alpha=p/q$ ($p$ and $q$ are co-prime
integers), the system has a unit cell of $q$ sites and the
single-particle spectrum is split into $q$ bands.
Given that the number of $\sigma$-component atoms is $N_\sigma$
and the number of lattice sites is $L$, we define the
component-depending band filling factor as
$\nu_{\sigma}=N_{\sigma}/N_{cell}$ with $N_{cell}=L / q$ being the
number of primitive cells. For a Fermi system, the system with the
band filling factor $\nu_\sigma = m$ ($m$ is an integer smaller
than q) corresponds to a band insulator with the lowest $m$ bands
being fully filled by the $\sigma$-component fermion. Such a band
insulator has been demonstrated to be characterized by a
nontrivial topological Chern number
in a 2D parameter space spanned by momentum and the phase of $\delta$ \cite{Lang}.
In this work
we shall consider the two-component system
with fractional filling factors $\nu_\uparrow$ and
$\nu_\downarrow$ but the total band filling factor
$\nu=\nu_\uparrow+\nu_\downarrow$ being an integer. For the noninteracting system,
the sub-band is
only partially filled and the system is a topologically trivial
conductor. We shall show that a Mott phase can emerge from the
conducting phase with an integer total filling factor when the
interaction effect is considered.

\subsection{ Emergence of Mott phase}
To give a concrete example, we first consider the FF mixture
described by the Hamiltonian (\ref{H0}) and (\ref{HI}) with
$U_\sigma=0$. We consider the case with $\alpha=1/3$ and $\nu=1$,
for which $n=N/L=1/3$, and calculate the charge gap defined as
$\Delta = [E_0(N+1) + E_0(N-1)]/2 - E_0(N) $ by numerically
diagonalizing the Hamiltonian, where $E_0(N)$ represent the ground
state (GS) energy for the N-atom system. In the Fig.1a, we show the charge gap
versus $U$ for the equal-mixing case with
$\nu_\uparrow=\nu_\downarrow=1/2$. For the system with either $N=4$ or $N=6$, the charge gap
increases with increasing $U$ and tends to $\Delta_b/2$ in the
large U limit,
where $\Delta_b$ represents the band gap between the lowest band
and the second one. The numerical results indicate that a Mott
insulator is emergent in the strongly interacting limit as the gap
is induced by the interaction.

Next we calculate the charge gap versus  $U$ for various $U'$ for
the equal-mixing FB mixture. Here we use $U'$ to represent
intra-component bosonic interaction strength for the
two-component FB mixture. As shown in Fig.1b for the system of $N=4$, with increasing
$U'$, the charge gap increases and the curve approaches the curve
of FF mixture. In the limit of hard-core boson, i.e., $U_\sigma
\rightarrow \infty$ for the BB mixture and $U'
\rightarrow \infty$ for the FB mixture, both the BB and FB model
can be mapped into the Fermi Hubbard model by extended Jordan-Wigner
transformations \cite{Chen_EPL}. Therefore, the above discussion
on the FF mixture can be directly applied to hard-core-boson limit
of the FB and BB mixtures.

\begin{figure}[tbp]
\includegraphics[scale=0.46]{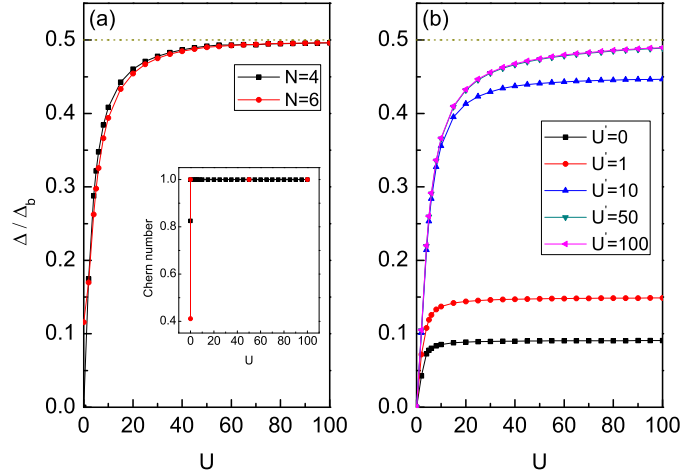}
\caption{(Color online) The charge gap versus the interaction
strength U for equal-mixing Fermi-Fermi system (a) and Fermi-Bose
system (b) with
 $\alpha=1/3$, $\lambda=1.5$, $\delta=0$,
and $n=1/3$ under periodic
boundary conditions (PBC). The inset in (a) shows the Chern number versus U.} \label{Fig1}
\end{figure}

\subsection{ Atomic mixtures in the strongly interacting limit}
The emergence of the charge gap for the superlattice system with
integer band filling factor can be clearly understood in the
limit of ISI, where we can construct the many-body wavefunction of the FF
system exactly by using the hard-core contact boundary condition (HCCBC)
\cite{Girardeau} and group theoretical methods \cite{Guan}.
Combining with the Pauli exclusion principle, the effect of an ISI
can be reduced to the HCCBC $ \Psi
\left( x_1,\sigma _1;\cdots ;x_N,\sigma _N\right) \mid
_{x_i=x_j}=0$, which does not depend on spin configurations.
According to Ref. \cite{Guan}, the many-body wave function can be
represented as $\Psi = \Psi_A \Psi_S$, where the spatial wave
function $\Psi_A$ is composed of Slater determinant of
$N=N_{\uparrow }+N_{\downarrow }$ orbitals $\phi _1(x),\cdots
,\phi _N(x)$, given by
\begin{equation}
\psi _A(x_1,\ldots ,x_N)=(N!)^{-\frac 12}det[\phi _n
(x_i)]_{i=1,\ldots ,N}^{n =1,\ldots ,N}  \label{psiA}
\end{equation}
with $\phi_n (x)$ ($x=ia$ with $a$ the lattice constant) the
eigenstate of the single particle Hamiltonian $H_0$, whereas
$\Psi_S$ is a mapping function composed of linear combination of
production of sign function and basis tensor of spin, whose
explicit form is given in Ref. \cite{Guan}. Effectively, the
effect of infinite repulsion is to generate a Pauli exclusion
between different components of fermions,
which has been experimentally observed in a two-particle system of fermionic
$^6$Li atoms with tunable interactions \cite{Zurn}.
\begin{figure}[tbp]
\includegraphics[scale=0.5]{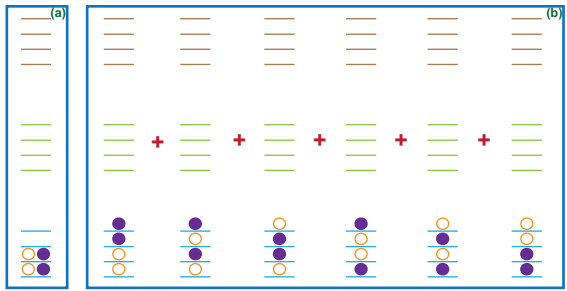}
\caption{(Color online) Schematic diagram for the ground state
energy level occupations for system with $\alpha=1/3$,
$N_{\uparrow}=N_{\downarrow}=2$ and $L=12$. (a) is for $U = 0$ and
(b) is for $U \to \infty$.}\label{Fig2}
\end{figure}

Now we can understand the formation of Mott phase by considering
the case in the limit of ISI. As illustrated in Fig.2, for the
superlattice with $\alpha=1/3$, the energy levels split into three
bands. When $U=0$, the lowest band is only partially filled for
$\nu_\uparrow = \nu_\downarrow = 1/2$. However the lowest band is
fully filled in the limit of $U \rightarrow \infty$ corresponding
to $\nu=1$ as each energy level can only be occupied by a fermion
with either spin up or spin down, and consequently a charge gap is
opened. According to the definition of charge gap, we have $\Delta
=\Delta_b/2 $ as adding a fermion costs the energy of $\Delta_b$,
which is consistent with the numerical result displayed in Fig.1.
For a general case with $\alpha=1/q$, the energy levels are
composed of q bands. As long as $\nu = m$ with $m=1, \cdots, q-1$,
a Mott insulator is formed as $U \rightarrow \infty$. For a
fractional $\nu$, for example, $\nu<1$, the lowest band is not
fully filled even in the limit of ISI, and thus no finite charge
gap is opened.

The above discussion can be directly applied to the BB mixture and
FB mixture in the limit of ISI. As $U \rightarrow \infty$ and
$U_{\sigma} \rightarrow \infty$, the effect of ISIs can be also
reduced to the HCCBC, which
enforces an effective Pauli exclusion to hard-core bosons and
between different components of atoms. Consequently, the exact
wave function of the system can be also represented as $\Psi =
\Psi_A \Psi_S$, where $\Psi_A$ is identical to the expression of
(\ref{psiA}), but $\Psi_S$ has different form for different kind
of mixtures \cite{Guan,Deuretzbacher,Girardeau}. For the BB
mixture, $\Psi_S = \prod_{1 \leq i,j \leq N} sgn(x_i-x_j)$, where
$sgn(x)$ is the sign function. The explicit form of the mapping
function $\Psi_S$ for the FB mixture is given in Ref.
\cite{Girardeau}.
\begin{figure}[tbp]
\includegraphics[scale=0.46]{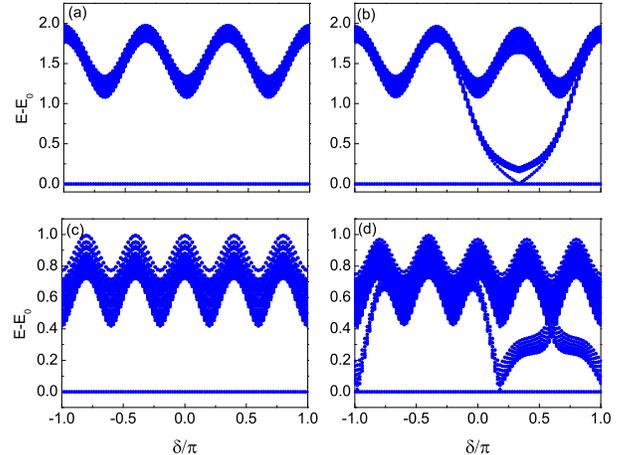}
\caption{(Color online) Low-energy spectrum versus $\delta$ for
two-component mixture systems with $\lambda=1.5$ and different
$\alpha$ in the infinitely repulsive limit. (a) $\alpha=1/3$,
$N=30$, $L=90$ under PBC; (b) $\alpha=1/3$, $N=30$, $L=91$ under
OBC; (c) $\alpha=1/5$, $N=40$, $L=100$ under PBC; (d)
$\alpha=1/5$, $N=40$, $L=101$ under OBC. }\label{Fig3}
\end{figure}

\subsection{ Topological feature of the Mott phase}
To characterize the
topological feature of the many-body states, it is convenient to
introduce a generalized boundary phase $\theta$ by applying the
twist boundary condition (TBC).
In the 2D parameter space of $(\theta,\delta)$, we can calculate
the Chern number of the many-body state, which is defined as an
integral invariant $C=\frac{1}{2\pi}\int{d\theta d\delta
F(\theta,\delta)}$, where $F(\theta,\delta) = \mathrm{Im}(\langle
\frac{\partial \Psi}{\partial \delta}| \frac{\partial
\Psi}{\partial\theta}\rangle -\langle \frac{\partial
\Psi}{\partial \theta}| \frac{\partial\Psi}{\partial\delta}
\rangle)$ is the Berry curvature \cite{TKNN,Niu}. For the
two-component mixtures in the limit of ISI, we notice that the
many-body wave functions under TBC can be
represented as $\Psi(\theta,\delta) = \Psi_A (\theta, \delta)
\Psi_S$, where only $\Psi_A$ varies with the change of $\theta$
and $\delta$, whereas $\Psi_S$ is independent of $\theta$ and
$\delta$ as it is composed of combination of production of sign
functions. This greatly simplifies the calculation of the Chern
number as $F(\theta,\delta) = \mathrm{Im}(\langle \frac{\partial
\Psi_A}{\partial \delta}| \frac{\partial
\Psi_A}{\partial\theta}\rangle -\langle \frac{\partial
\Psi_A}{\partial \theta}| \frac{\partial\Psi_A}{\partial\delta}
\rangle)$, where $\Psi_A$ is given by Eq.(\ref{psiA}).
Consequently,
the Chern number for atomic mixtures composed of $N_\uparrow$ and $N_\downarrow$
two-component atoms is identical to the Chern number of the system
composed of $N$ free fermions. For the system with $\alpha=1/3$,
the Mott insulators are formed when $\nu=1$ and $\nu=2$,
corresponding to states with Chern number $C=1$ and $C=-1$, which
characterizes the Mott insulators being topologically nontrivial.
For the system with a finite $U$, the Chern number can be obtained
via numerically calculating the many-body wavefunction under
TBC \cite{Niu}. As shown in the inset of Fig.\ref{Fig1}a,
the Chern number for systems with $\nu=1$ approaches 1 even for a very small $U$.
Such a result reminds us a similar phenomenon in the 1D Hubbard
model without periodic modulation, for which the half-filling Hubbard system
enters the Mott phase for an arbitrary repulsive interaction as shown in
the seminal work of Lieb and Wu \cite{Lieb-Wu}. The results for finite $U$ indicate that our exact
conclusions in the limit of ISI are robust even when the interaction deviates the limit of ISI.
When the filling deviates $\nu=1$, the Chern number is no longer an integer number.

According to the bulk-edge correspondence in general topological
insulators, one may expect that the TMIs should display nontrivial
edge states for the system with open boundary conditions (OBC). To
see it clearly, we display the excitation spectrum versus the
phase $\delta$ for mixture systems in the strongly interacting
limit with $\nu=1$, $\alpha=1/3$ and $\nu=2$, $\alpha=1/5$ in
Fig.\ref{Fig3}. The system with $\nu=1$ and $\alpha=1/3$ is a TMI
characterized by the Chern number $C=1$, whereas the TMI
corresponding to $\nu=2$ and $\alpha=1/5$ is characterized by
$C=2$. As shown in Fig. \ref{Fig3}a and Fig. \ref{Fig3}c, there is
an obvious gap between the GS and the first excited
state for systems with PBC. However, as shown in Fig. \ref{Fig3}b
and Fig. \ref{Fig3}d, edge states appear in the gap regimes for
systems with OBC. As the phase varies from $-\pi$ to $\pi$, the
edge states connect the GS to the excited band.

\begin{figure}[tbp]
\includegraphics[scale=0.46]{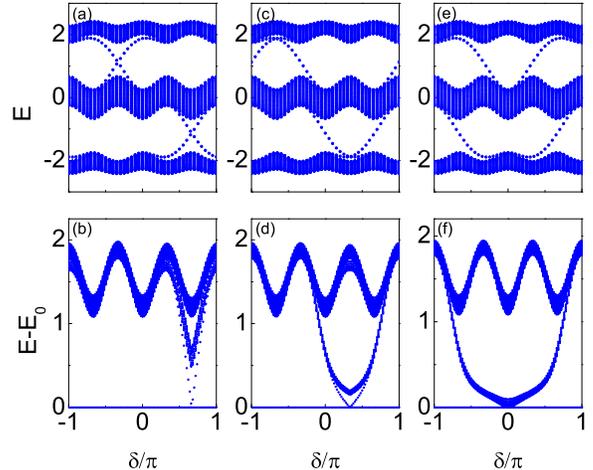}
\caption{(Color online) Low-energy spectrum versus $\delta$ for
the systems with $\lambda=1.5$, $\alpha=1/3$, and various $L$
under open boundary condition. The top three figures are for single-particle
spectrums and the bottom ones are for many-body systems with $N=30$ in
the strongly repulsive limit. Three columns from left to right are with $L=90$, $91$
and $92$, respectively.}\label{sfig1}
\end{figure}

For the periodic modulation system under PBC considered in the current work, the unit cell is composed of $q$ different sites, and one need take the lattice size $L$ as $L=q N_{cell}$ to fulfill the PBC. However, for the system under OBC, one can take the lattice size $L= q N_{cell}+i$ with $i=0,\cdots,q-1$, i.e., we have $q$ different choices \cite{note}. For example, for the periodic system with $\alpha=1/3$ and $L=90$, we should have three different choices of lattice sizes with $L=90$, $91$, and $92$ under OBC. In Fig. \ref{Fig3}b, we only present the example with $L=91$. To see clearly the effect of lattice sizes,
here we present the energy spectrums versus $\delta$ with different lattice sizes under OBC in Fig.\ref{sfig1}. The top three figures in Fig.\ref{sfig1} are for single-particle spectra and the bottom ones
are for many-body systems of $N=30$ in the strongly repulsive limit. Three columns from left to right are with $L=90$, $91$
and $92$, respectively. As shown in the figure,
the positions of the gapless single-particle
edge modes change with the change of lattice sizes, and correspondingly shapes of many-body edge modes also change. However, the edge modes always connect ground state and excited bands in the bulk gap regime for different sizes.
It is clear that the choice of different lattice sizes under OBC affects the concrete shape of edge modes,
but does not change its topological properties demonstrating by edge modes connecting ground state and excited bands in the band gap regime of the corresponding bulk system.

\begin{figure}[tbp]
\includegraphics[scale=0.46]{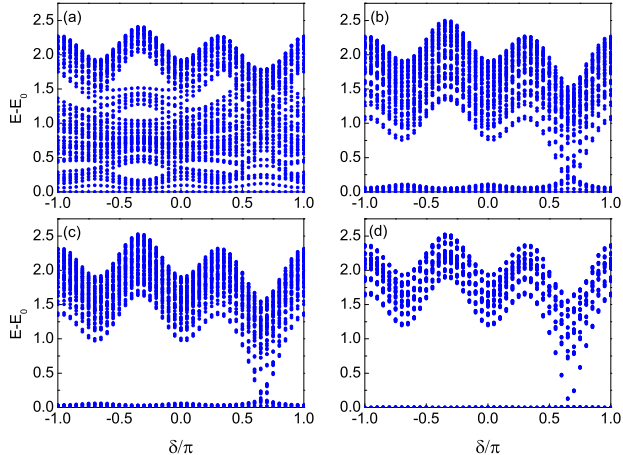}
\caption{(Color online) Low-energy spectrum versus $\delta$ for
the Fermi-Fermi mixture with $\alpha=1/3$, $\lambda=1.5$
$N_{\uparrow}=N_{\downarrow}=2$, $L=12$ and various $U$ under open boundary condition.
(a) $U=1$, (b) $U=5$, (c) $U=10$ and (d) $U=100$.} \label{sfig2}
\end{figure}

\begin{figure}[tbp]
\includegraphics[scale=0.46]{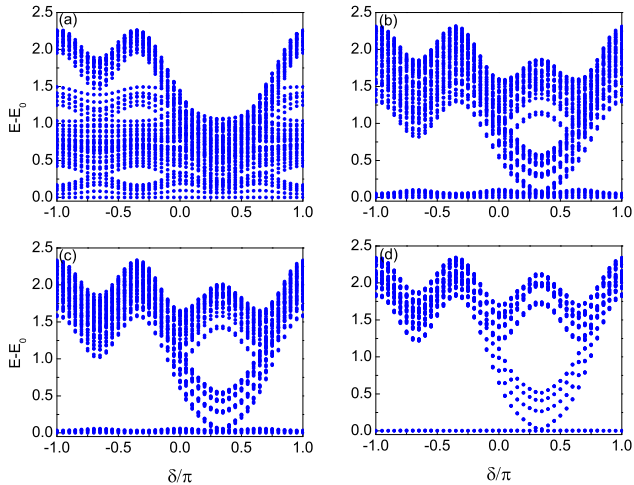}
\caption{(Color online) Low-energy spectrum versus $\delta$ for
the Fermi-Fermi mixture with $\alpha=1/3$, $\lambda=1.5$
$N_{\uparrow}=N_{\downarrow}=2$, $L=13$ and various $U$ under OBC.
(a) $U=1$, (b) $U=5$, (c) $U=10$ and (d) $U=100$.} \label{fig6}
\end{figure}

When the interaction deviates from the limit of ISI, despite of
the exact mapping no longer holding true, our numerical results
indicate that the topological Mott phase still exits. To see how
the edge states change when $U$ deviates the limit of ISI, we display the excitation spectrum versus the
phase $\delta$ for the FF mixture system with $\alpha=1/3$, $N_{\uparrow}=N_{\downarrow}=2$, $L=12$ and various $U$ under the OBC in
Fig.\ref{sfig2}. As shown in
Fig.\ref{sfig2}, the excitation spectrum for the system with
$U=100$ exhibits almost the same behavior of system in the limit of ISI as shown in Fig.4b. For $U=10$ and $U=5$,
the spectra
still have similar structures but with the
GS levels being broadened for smaller $U$ due to spin fluctuations as the
degeneracy of GSs in the limit of ISI is lifted for
finite interactions. However, for $U=1$, no obvious edge modes can be detected as no an obvious gap regime
appears in the weakly interacting regime even under the PBC.
In order to see the effect of lattice size, we also present results for the system with $\alpha=1/3$, $N_{\uparrow}=N_{\downarrow}=2$ and a different lattice size of $L=13$ in Fig.\ref{fig6}. It is clear that the low-energy excitation spectra have similar behaviors as that of the system with $L=12$ shown in Fig.\ref{sfig2}. For both systems, the edge modes connect the lower energy parts and the higher excited ones, which is consistent with the results in the limit of ISI as shown in Fig.4(b) and Fig.4(d).


Our results can be directly extended to the general
multi-component atomic systems trapped in 1D superlattices.
When the inter-specie and inner-specie interactions between atoms go to
the strongly interacting limit, an effective Pauli exclusion
between atoms arises. Therefore, a TMI is expected to appear in
the topologically nontrivial superlattice as long as the total
filling factor $\nu=\sum_{\kappa} \nu_{\kappa}$ is an integer,
where $\kappa$ is the component index of the multi-component
atomic mixture.
\begin{figure}[tbp]
\includegraphics[scale=0.45]{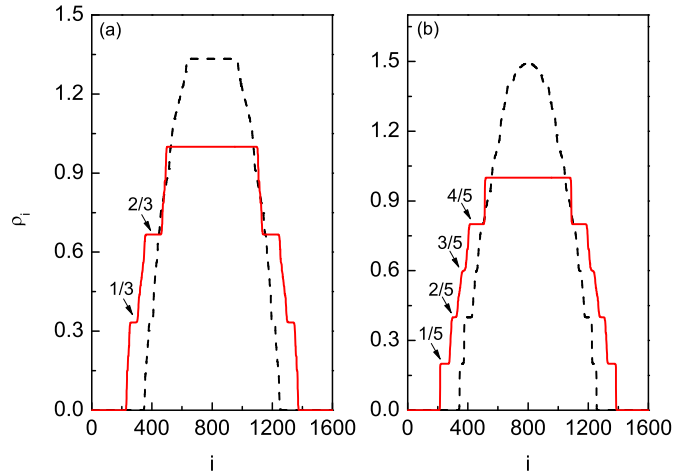}
\caption{(Color online) The local average density distributions
for the superlattice with $\alpha=1/3$ (a), and $\alpha=1/5$ (b)
trapped in the harmonic trap.  The system is with 1600 sites,
$N_{\uparrow}=400$, $N_{\downarrow}=500$, $\lambda=1.5$,
$\delta=0$ and $V_{H}=2\times10^{-6}$. The dash line is for the
case of $U=0$ and the solid line  is for the case of $U\to
\infty$. Here, we take $M=(q-1)/2$. }\label{Fig7}
\end{figure}

\subsection{Experimental detection}
A possible way to observe the TMI is to detect the density distribution of the superlattice systems
with an external confining potential, i.e., $V_i$ in Eq.(\ref{H0})
is replaced by $ V_i= \lambda \mathrm{cos}(2\pi \alpha
i+\delta)+V_H(i-i_0)^2$, where $V_H$ is the strength of the
harmonic trap with $i_0$ being the position of trap center.  In
Fig.7, we show the local average density distribution of
two-component fermions with imbalance populations trapped in the
optical superlattice with a harmonic trap. The local average
density is defines as $\rho_i=\sum_{j=-M}^M n_{i+j}/(2M+1)$ with
$M\ll L$ \cite{Lang}, where $n_i= \langle
\hat{n}_{i,\uparrow}\rangle + \langle \hat{n}_{i,\downarrow}
\rangle$. In the strongly interacting limit, two plateaus with
$\rho=1/3$ and $\rho=2/3$ appear for $\alpha=1/3$, whereas four
plateaus with $\rho=1/5,2/5,3/5,4/5$ appear for $\alpha=1/5$. We
note that locations of plateaus appear at $\rho (\alpha)=\alpha,
1-\alpha, 2\alpha, 1-2\alpha,...$, if the
values are in the range of $(0,1)$. For contrast, the density
profiles for $U=0$ do not exhibit these plateaus. The Chern number
can thus be deduced from the plateau distribution by using the
Streda formula \cite{Streda,Lang} $C=\frac{\partial \rho
(\alpha)}{\partial \alpha}$. It is straightforward to get $C=1,-1$
for $\rho_i=\alpha, 1-\alpha$ and $C=2,-2$ for $\rho_i=2\alpha,
1-2\alpha$. Alternative methods of detecting Chern numbers in
optical lattices through time-of-flight images have also been
proposed \cite{Zhao,WangL}.

\section{ Summary}
In summary, we studied the realization of TMIs in
interacting atomic mixtures trapped in 1D optical superlattices.
We give a clear interpretation for the formation of TMI by using
an exact mapping, which relates the many-body wavefunction of
atomic mixtures in the strongly interacting limit to that of the
free fermion system. The TMI displays nontrivial edge states and
can be characterized by a nonzero Chern number. The topological
feature of the Mott insulator can be revealed by detecting
plateaus of density profiles of the trapped lattice systems. Our
results pave the way for experimentally studying TMIs in 1D
optical lattices and can be directly extended to the general
multi-component atomic systems trapped in 1D superlattices.

\begin{acknowledgments}
We thank L.-J. Lang for helpful discussions. This work has been supported by National Program for Basic Research of MOST, NSF
of China under Grants No.11121063 and No.11174360, and 973 grant.
\end{acknowledgments}

\appendix
\section{}
In the appendix, we make a comparison between the interacting Fermi-Fermi mixtures described by Hamiltonian (1)
and (2) and the interacting Hofstadter
model described by $H^{2D}=H_0^{2D} + H_I^{2D}$ \cite{Doh}.
For a two-dimensional square lattice in a uniform perpendicular magnetic field $B$, under the Landau gauge $\vec{A}=B(0,x,0)$, the noninteracting part $H_{0}^{2D}$ is given by \cite{Hofstadter}
\begin{eqnarray}
H_{0}^{2D}&=& \sum_{i,j,\sigma=\uparrow,\downarrow}
-t_x\hat{c}^\dagger_{i,j,\sigma} \hat{c}_{i+1,j,\sigma} \nonumber\\
& &-t_y e^{-i2\pi\alpha i}\hat{c}^\dagger_{i,j,\sigma} \hat{c}_{i,j+1,\sigma} +\mathrm{H.c.},
\label{H2D0}
\end{eqnarray}
where $t_x$ and $t_y$ are the strength of the nearest-neighbor hopping along the $x$ and $y$ directions,
$\hat{c}_{i,j,\sigma}$ is the annihilation operator of the fermions with spin $\sigma$ at $(i,j)$ in the lattice,
and $\alpha$ is the ratio of flux through a unit cell to one flux quantum. For a Fermi-Fermi mixture, the on-site term at $(i,j)$ site in the lattices is given by
\begin{equation}
H_I^{2D} = U \sum_{i,j} \hat{n}_{i,j,\uparrow} \hat{n}_{i,j,\downarrow},
\label{H2DI}
\end{equation}
Taking a Fourier transformation in the $y$ direction
$\hat{c}_{i,j,\sigma}=\frac{1}{\sqrt{L_{k_y}}}\sum_{k_y}e^{-ik_y j}\hat{c}_{i,k_y,\sigma}$,
we can rewrite the Hamiltonian $H^{2D}=H_0^{2D}+H_I^{2D}$ in the $k_{y}$-momentum space as
\begin{eqnarray}
H_0^{2D} &=&\sum_{i,k_y,\sigma=\uparrow,\downarrow}
-t_x\left( \hat{c}^\dagger_{i,k_y,\sigma} \hat{c}_{i+1,k_y,\sigma}+\mathrm{H.c.} \right) \nonumber\\
& &-2t_y \cos(2\pi\alpha i+k_y)\hat{n}_{i,k_y,\sigma}
\end{eqnarray}
and
\begin{eqnarray}
H_I^{2D} &=& \frac{U}{L_y}\sum_i\sum_{k_{y_1},k_{y_2},k_{y_3},k_{y_4}}\delta_{k_{y_1}-k_{y_2}+k_{y_3}-k_{y_4},2\pi} \nonumber \\
& &\hat{c}^\dagger_{i,k_{y_1},\uparrow}\hat{c}^\dagger_{i,k_{y_3},\downarrow}\hat{c}_{i,k_{y_4},\downarrow}\hat{c}_{i,k_{y_2},\uparrow},
\label{H2Dky}
\end{eqnarray}
In the limit of $U \to 0$,  by making substitutions of $t_x \to t$, $-2t_y \to \lambda$, and $k_y \to \delta$,
the Hamiltonian $H^{2D}$ can be mapped to the one-dimensional model with a periodic or quasi-periodic modulation
(Hamiltonian (1) in the main text) which depends on $\alpha$ being rational or irrational number. In spite of the existence of such a mapping for the noninteracting systems, the interacting term $H_I^{2D}$ under the Fourier transformation displays obvious different form as the on-site
term of the Hamiltonian (2) in the main text. While the interacting term for the 1D model is only relevant to a given parameter of $\delta$,
the interacting term $H_I^{2D}$ couples different $k_y$ modes (corresponding to $\delta$) together. Therefore, the exact results for the one-dimensional systems in the strongly interacting limit can not be applied to the corresponding two-dimensional interacting Hofstadter model directly.


\begin{references}
\bibitem{Supp1} We note that the mathematical
formulation of the current problem reduces to the Harper-Hofstadter model when $U=0$, however the general problem with
many-body interactions included can not be mapped to the 2D interacting Hofstadter problem as considered in Ref.\cite{Doh}.
See the appendix for details.
\bibitem{note} For the superlattice system with $\alpha=1/q$ under OBC, we note that behaviors of edge modes
for the system with size of $L+q$ are almost the same as that of the system with size of $L$.
\end{references}
\end{document}